\documentclass[doublecol]{epl2}

\begin{document}

\title{The game of go as a complex network}

\author{B. Georgeot\inst{1,2}\and O. Giraud\inst{3}} 
\shortauthor{B. Georgeot and O. Giraud} 
\institute{
 \inst{1} Universit\'e de Toulouse; UPS; Laboratoire de
 Physique Th\'eorique (IRSAMC); F-31062 Toulouse, France\\
\inst{2} CNRS; LPT (IRSAMC); F-31062 Toulouse, France\\
\inst{3} Univ.~Paris-Sud,  CNRS, LPTMS, UMR 8626, Orsay, F-91405, France
}
\date{\today} 

\abstract{
We study the game of go from a complex network perspective.
We construct a directed network using a suitable definition of
tactical moves including local patterns, and study this network for
different datasets of professional and amateur games.
The move distribution follows Zipf's law and the network is scale free, with statistical peculiarities different from other real directed networks,
such as e.~g.~the World Wide Web. These specificities reflect in the outcome of ranking algorithms applied to it.  The fine study of the eigenvalues and eigenvectors of matrices used by the ranking algorithms 
singles out certain strategic situations.
Our results should pave the way to a better modelization of board
games and other types of human strategic scheming.
}

\date{May 12, 2011}

\maketitle

%%%%%%%%%%%%%%%%%%%%%%%%%%%%%%%%%%%%%%%%%%%%%%%%%%%%%%%%%%%%%%%%%%%%%%%%%%%%%%%
%******************************************************************************
\section{Introduction}
%******************************************************************************
%%%%%%%%%%%%%%%%%%%%%%%%%%%%%%%%%%%%%%%%%%%%%%%%%%%%%%%%%%%%%%%%%%%%%%%%%%%%%%%
The study of complex networks has attracted increasing interest in the
past decade, fueled in particular by the great recent development 
of communication and information networks.  
Tools and models have been created, enabling to describe the growth 
mechanisms and properties of such systems.  In parallel, it has been realized
that many important aspects of the physical world or of social interactions
can be modelized by such networks.  Such tools have been applied to many fields of human activity, such as e.g. languages or friendships
\cite{barabasi}.

One of the oldest activities of human beings, board games have been played for millenia. Besides their intrinsic interest, they represent a privileged approach to the working of decision-making in the human brain. Some of them are very difficult to
modelize or simulate: only recently were computer programs able to beat world chess champions. The old Asian game of go is even less tractable. The game complexity, that is, the total number of legal positions, is about $10^{171}$, compared to a mere $10^{50}$ for chess \cite{TroFar07}. It remains an open challenge for computer scientists: while Deep Blue famously beat the world chess champion Kasparov in 1997, no computer program has beaten a very good player even in recent times. 

Many studies have been devoted to ``computer go'', the simulation of the go game on a computer. They were historically based on deterministic tree search algorithms such as minimax or alpha-beta, which estimate an evaluation function (giving the game-theoretic value of a move) on the tree of allowed moves (see e.~g.~\cite{computers,BouCaz01}). Recently, much progress has been done by introducing Monte-Carlo techniques \cite{MonteCarloGo, progressive}, which basically estimate the value of a move by playing subsequent moves at random according to the rules of go. Monte-Carlo tree search algorithms are based on a non-uniform probability distribution over legal moves, and explore only the most promising ones. Variations on these techniques allowed computer programs to make significant progress in the last few years, so that professional human players with a large enough handicap could be beaten by a computer \cite{LeeWang09}. The choice of the evaluation function and the way in which the tree is explored are crucial ingredients for any further progress. 

Since go is a popular game with millions of players in the world, many games have been recorded, which enables statistical data to be extracted reliably.
%These databases of real games can be used to feed strategic information into the program simulating go, such as the frequency at which a given move is played in a given local situation.  In this case random moves can be replaced by pseudo-random moves, allowing a proper balance between exploitation of successful moves and exploration of original moves.  
A few works have used statistical properties of recorded professional games to optimize performances of Monte-Carlo algorithms. 
Usually the simplest features of real games are retained, such as local patterns or contiguity to the previous move \cite{Coulom}; including more real-game features improves noticeably the winning rate of computer programs \cite{Coulom2}. 

In this paper, we thus study the game of go from a complex network perspective. We use databases of expert games in order to construct networks from the different sequences of moves, and study the properties of these networks.  We based our numerical results on the whole available record, from 1941 onwards, of the most important historical professional Japanese go tournaments: Kisei (143 games), Meijin (259 games), Honinbo (305 games), Judan (158 games) \cite{database}.  To increase statistics and compare with professional tournaments, 4000 amateur games also available from \cite{database} were used. 
%While other approaches based on statistical physics are possible, the complex network perspective seems to us to be the closest to the spirit of works in computer science. We think that the potential knowledge acquired from the powerful tools of complex networks applied to the game of go would be profitably taken into account in future go programs.

%%%%%%%%%%%%%%%%%%%%%%%%%%%%%%%%%%%%%%%%%%%%%%%%%%%%%%%%%%%%%%%%%%%%%%%%%%%%%%%
%******************************************************************************
\section{Definition of inequivalent moves}
%******************************************************************************
%%%%%%%%%%%%%%%%%%%%%%%%%%%%%%%%%%%%%%%%%%%%%%%%%%%%%%%%%%%%%%%%%%%%%%%%%%%%%%%
The game of go is played by two players (Black and White) on a board (goban) consisting of 19 horizontal and 19 vertical lines. The players alternately place a stone of their own color at an empty intersection on the board. Stones entirely surrounded by the opponent must be removed, and the aim of the game is to delimit large territories.  As the game unfolds, local and global properties of stones are involved. A network approach will obviously not be able to capture all features of the game, as the number of possible moves is far too large. Here we follow an approach where only local features are retained. This approach is reminiscent of the one used in the context of language networks \cite{language}. 

A move consists in placing a stone at an empty intersection $(h,v)$ with $1\leq h,v\leq 19$. We call ''plaquette'' a square of $3\times 3$ intersections, that is, a subset of the board of the form  $\{(h+r,v+s),-1\leq r,s\leq 1\}$ (to account for edges and corners of the board one can imagine that there are two additional dummy lines at each side of the board). To define our network we only take into account intersections closest to $(h,v)$. Vertices correspond to the different kinds of plaquettes in which a player can put a stone, irrespective of where it has been played on the board.
Since each of the 8 neighboring intersections can be either empty, black or white, there are $\sim 3^8$ different plaquettes. We choose to consider identical plaquettes that transform to each other under any symmetry of the square (rotation or flip). We also identify patterns with color swapped. That is, a move where Black plays in a given plaquette is considered the same as a move where White plays in the same plaquette with colors swapped. An exact computation taking into account borders and symmetries leaves us with 1107 nonequivalent plaquettes with empty centers, which are the vertices of our network. We note that certain computer programs based on knowledge from real professional games also consider similar $3\times 3$ stone patterns \cite{Coulom, Coulom2}. Considering larger plaquettes is possible and would convey more relevant information; however, the number of vertices then becomes enormously large ($\approx 3.10^{10}$ for $5 \times 5$ plaquettes). 

\begin{figure}[h]
\begin{center}
\includegraphics*[width=7.5cm]{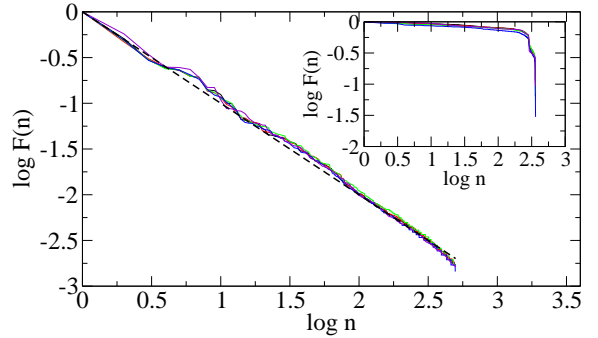}
\end{center}
\caption{(Color online) Normalized integrated frequency distribution of moves $F(n)$ for Honinbo (black), Meijin (red), Judan (green), Kisei (blue) and amateur (violet) tournaments. The normalized number of occurrences of the 500 most frequent moves (among the 1107 moves described in the text) is shown vs the ranks of the moves (rankings slightly depend on the database). Slopes are resp. -1.058, -1.056, -1.065, -1.067, -1.081. Thick dashed line is $y=-x$.
Inset: same with moves defined as position of the stone on the board. Log. is decimal.
\label{zipf_all}}
\end{figure}

This definition of inequivalent moves enables us to
investigate the first properties of the databases in term of frequencies 
of moves.
Zipf's law is an empirical characteristics which has been observed in many natural distributions, such as e.~g.~word frequency in the English language~\cite{zipf}, city sizes~\cite{zipf1}, income distribution of companies~\cite{zipf3}, and chess openings~\cite{chess}.
%It was first observed in the frequency distribution of 
If items are ranked according to their frequency, it predicts a power-law decay of the frequency as a function of the rank. 
%Namely, the $n$th word by order of frequency has a frequency which scales as $\propto 1/n^2$, or equivalently the integrated frequency distribution scales as $F(n)\propto 1/n$. 
Zipf's law was observed in the frequency distribution of $5\times 5$ go patterns \cite{LiuDouLu08}. In Fig.~\ref{zipf_all} we display the integrated frequency distribution for our 1107 moves labeled from the most to the least frequent. The integrated distribution of moves is very similar for all databases and clearly follows a Zipf's law, with an exponent $\approx 1.06$.
%close to the one observed for languages or cities. 
In contrast, such a law cannot be seen if one simply takes the $361$ possible positions $(h,v)$ 
as vertices, disregarding local features (inset of Fig.~\ref{zipf_all}). Thus Zipf's law appears when tactical information is taken into account. For all databases the 10 most frequent moves are the same (see Fig.~\ref{frequent_moves}, upper line), but sometimes in a slightly different order.

%%%%%%%%%%%%%%%%%%%%%%%%%%%%%%%%%%%%%%%%%%%%%%%%%%%%%%%%%%%%%%%%%%%%%%%%%%%%%%%
%******************************************************************************
\section{The go network}
%******************************************************************************
%%%%%%%%%%%%%%%%%%%%%%%%%%%%%%%%%%%%%%%%%%%%%%%%%%%%%%%%%%%%%%%%%%%%%%%%%%%%%%%

\begin{figure}[h]
\begin{center}
\includegraphics*[width=7.2cm]{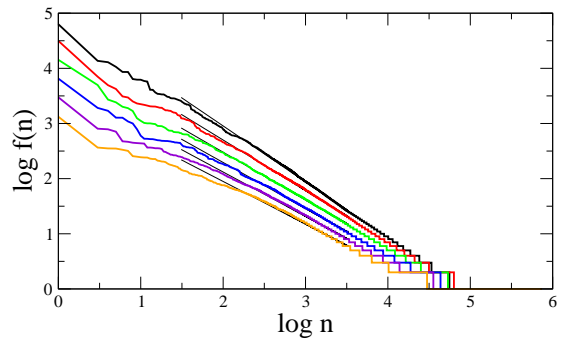}
\end{center}
\caption{{(Color online) Integrated frequency distribution of sequences of moves $f(n)$ for (from top to
bottom) two to
seven successive moves (all databases together), plotted against the ranks of the moves. Moves are
the 1107 moves described in the text. Slopes are 
resp. -1.01, -0.91, -0.86, -0.83, -0.81, -0.77.
\label{frequencies1} }}
\end{figure}

\begin{figure}[h]
\begin{center}
\includegraphics*[width=7.2cm]{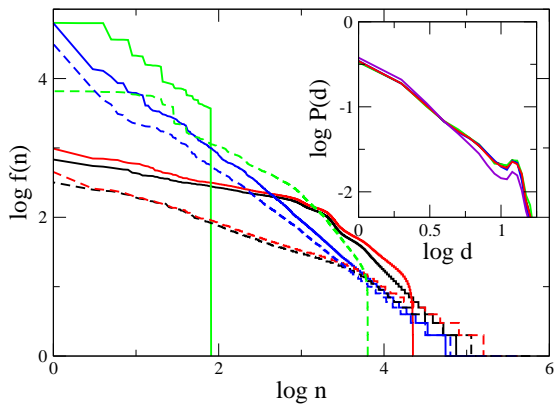}
\end{center}
\caption{{(Color online) Integrated frequency distribution of sequences of moves $f(n)$ for sequences of two (continuous lines) and three (dashed
lines) successive moves for (from bottom to top) case C1 (black, slopes -0.23, -0.4), C2 (red, slopes -0.25, -0.39), same curves as in the main panel (blue), C3 (green, -0.91, -0.70). Inset: distribution of distances between moves $P(d)$; same color code as in Fig. 1. The four professional tournaments are almost undistiguinshable. \label{frequencies2} }}
\end{figure}

The dynamics of the game is built from successive moves. In the game of go, a game often consists in
a series of small fights played at different places on the board.
A player might put a stone in the vicinity of their
opponent's stones to engage the battle, but the opponent might prefer to
first continue a fight occurring somewhere else, in which case two 
consecutive moves would not be directly related. In order to construct our network, it is thus natural to connect 
two moves by a directed link only if these moves follow each other in the same region of the board. 
To be more specific, we connect vertices corresponding to moves $a$ and $b$ played at $(h_a,v_a)$ and $(h_b,v_b)$ respectively if $b$ follows $a$ in a game and $\max\{|h_b-h_a|,|v_b-v_a|\}\leq d$. Each choice of the integer $d$ defines a different network. The choice of $d$ determines the distance beyond which two moves are considered nonrelated. 
We present in Fig.~\ref{frequencies1} the frequency distribution for sequences of moves defined in such a way with $d=4$. We observe an algebraic decrease with exponent ranging from $\approx 1$ for short sequences to $\approx 0.8$ for longer sequences. Thus sequences of moves follow Zipf's law, as was observed for word sequences in languages \cite{language}.  We attribute the decrease of the exponent to the fact that longer sequences reflect more and more individual strategies. As a comparison, Fig.~\ref{frequencies2} displays the frequency distribution of successive moves for three other definitions of moves and sequences of moves. In cases C1 and C2, moves are defined as positions $(h,v)$ on the board, disregarding local features, and $b$ is considered to follow $a$ if $b$ is played immediately after $a$ (case C1) or if $b$ is the first move played after $a$ and in the vicinity of $a$,  with $d=4$ (case C2); in case C3 sequences of vectors between two moves played in the same region (with $d=4$) are considered. These results indicate that move sequences, even long ones, are best hierarchized by our initial definition. In what follows we will thus disregard other choices C1-C3. In the inset of Fig.~\ref{frequencies2}, we also show the distribution $P(d)$ of distances between consecutive moves. Interestingly enough, the amateur database departs significantly from all the professional ones, with a tendency to play more often at shorter distances. This may reflect the fact that professionals are more prone to play several tactical fights in parallel, or play on average shorter local tactical fights.

We now investigate the properties of our networks. We construct a network for each database by playing the games according to the rules of go and adding directed links between the 1107 vertices as indicated above. To each link is assigned a weight given by the number of occurrences in the database. 

\begin{figure}[h]
\begin{center}
\includegraphics*[width=7.6cm]{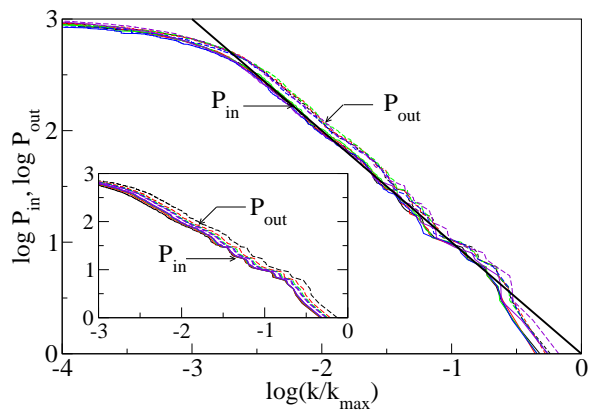}
\end{center}
\caption{(Color online) Normalized integrated distribution of ingoing links $P_{\textrm{in}}$ (lower curves, solid) and outgoing links $P_{\textrm{out}}$ (upper curves, dashed), for networks built with $d=4$. The number of vertices with more than $k$ ingoing (outgoing) links is shown vs the normalized number of ingoing (outgoing) links $k/k_{\textrm{max}}$. Same databases and color code as in Fig.~\ref{zipf_all}. Thick solid line is $y=-x$. Inset:  $P_{\textrm{in}}$ (solid curves) and $P_{\textrm{out}}$ (dashed curves) for the Honinbo database, $d=2$ (black), 3 (red), 4 (green), 5 (blue) and 6 (violet). \label{pinpout}}
\end{figure}

The distribution of ingoing and outgoing links $P_{\textrm{in}}$ and  $P_{\textrm{out}}$ is displayed in Fig.~\ref{pinpout}. The tails of both distributions are very close to a power-law $1/k^{\gamma}$ with exponent $\gamma=1.0$ for the integrated distribution.  The results are stable in the sense that
the exponent does not depend on the database considered.  The presence of such power laws indicates that the network displays the scale-free property: the distribution of links around a given link frequency is independent of that frequency. Such a property has been seen in many social or biological networks, but is absent in e.g. the famous Erd\"os-Renyi model of random networks.  The symmetry between ingoing and outgoing links is a peculiarity of this network; it is well known for the World Wide Web (WWW) for
example that the exponent for $P_{\textrm{out}}$ ($\gamma\approx 1.7$) is much larger
than for $P_{\textrm{in}}$ ($\gamma\approx 1.1$) \cite{donato}.  In the case of the WWW, the number of outgoing links is limited by the behavior of each independent webmaster.  In our case, the results indicate a symmetry, at least at a statistical level, between moves that often follow others and moves which
have many possible following moves. This may correspond to the fact that many short tactical sequences can be played in a different order within several different contexts.  In order to analyze the dependence of  $P_{\textrm{in}}$ and $P_{\textrm{out}}$ on the choice of $d$, we plot these quantities for a network constructed for various values of $d$ in the inset of Fig.~\ref{pinpout}. The distribution of ingoing and outgoing links stabilizes at $d=4$. Other databases give similar results (data not shown).  We now focus on $d=4$.

Some properties can be extracted from the unweighted adjacency matrix
(i.~e.~without weighing the links according to their frequency). The clustering coefficient (CC) describes the tendency of 
many real-world networks to form local clusters of highly connected vertices. 
%For undirected networks, 
The CC of a given vertex $i$ is defined as the
probability that two neighbors of $i$ be connected to each other, irrespective
of the direction of the link.
%Various
%extensions of this quantity have been proposed for directed and weighted
%networks. 
%Here we use the definition introduced in\cite{directed_clustering}. 
The average CC for our networks is displayed in Fig.~\ref{log_sorted_pageranks} (inset).  The CC depends on the number of games $n_g$ included to construct the network, but almost not on the database. For large $n_g$, the CC goes to
an asymptotic value which appears to be larger than $0.7$, indicating
high clustering, larger than the WWW (where the CC is $\approx 0.11$ \cite{barabasi}) and comparable to social \cite{barabasi} or language networks
\cite{language}.

%%%%%%%%%%%%%%%%%%%%%%%%%%%%%%%%%%%%%%%%%%%%%%%%%%%%%%%%%%%%%%%%%%%%%%%%%%%%%%%
%******************************************************************************
\section{Ranking vectors for the go network}
%******************************************************************************
%%%%%%%%%%%%%%%%%%%%%%%%%%%%%%%%%%%%%%%%%%%%%%%%%%%%%%%%%%%%%%%%%%%%%%%%%%%%%%%

\begin{figure}[h]
\begin{center}
\includegraphics*[width=7.5cm]{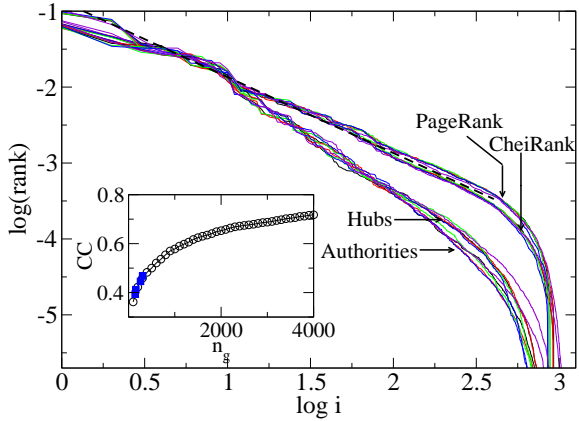}
\end{center}
\caption{(Color online) Ranking vectors for matrices $G$ with $\alpha=1$. Same color code as in Fig.~\ref{zipf_all}, $d=4$. From top to bottom, top bundle: PageRank.
%linear fit in the range $2\leq \ln(i)\leq 6$ gives slopes respectively -1.0761, -1.0776, -1.0877, -1.0934, -1.0938. 
Second bundle: CheiRank. Third bundle: Hubs. Fourth (bottom) bundle: Authorities. Straight dashed line is $y=-x$.
Inset: Clustering coefficient (CC) as a function of the number
of games $n_g$ included to construct the network; blue squares: professional tournaments, all games included; circles: amateur games. \label{log_sorted_pageranks}}
\end{figure}

In order to get an insight into how our network captures aspects of the 
strategy of the game, we now use the weighted adjacency matrix (links are weighted according to their frequency in the database) and apply tools developed
to rank vertices in order of importance, used e.g.~to determine
the order of appearance of answers to queries by search engines.  These
methods generally build a ranking vector, whose value on each vertex enables
to determine its importance. The most famous such vector is the PageRank vector
\cite{brin,googlebook}, which was the basis of the Google search engine. It
is built from the Google matrix $G$, defined as $G_{ij}=\alpha S_{ij}+(1-\alpha)\, ^{t}ee/N$, where $e=(1,...,1)$, $N=1107$, $0<\alpha\leq 1$, 
$S$ is the weighted adjacency matrix with any column of 0 replaced by a column of $1$, and the sum of each column normalized to 1. The PageRank vector is the right eigenvector of $G$ associated with the largest eigenvalue $\lambda=1$, and singles out vertices with many incoming links from important nodes. 
From its definition, its components are real and nonnegative, and therefore
can be used to rank nodes according to the value of these components.
Other ranking vectors built from $G$ include
the CheiRank vector \cite{dima} (which is the PageRank of the network with all links inverted), and the Hubs and Authorities of the HITS algorithm \cite{hits}. They all share the properties of being real nonnegative vectors, and thus can be used to rank the nodes of the network. While PageRanks and Hubs reflect properties of vertices depending on their incoming links,
CheiRanks and Authorities are based on outgoing links.
%which emphasizes moves which can be followed by many others, and the Hubs and Authorities of the HITS algorithm \cite{hits}, 
%which emphasize moves following many others (Authorities) 
%or moves with many possible followers (Hubs). 
In Fig.~\ref{log_sorted_pageranks} we show these ranking vectors for our networks.
They follow an algebraic law with slope $\approx -1$ (PageRank and CheiRank)
and $\approx -1.5$ (Hubs and Authorities).  A similar distribution of the PageRank
was observed in e.g. the WWW \cite{donato,google},
but in contrast with the WWW and other systems there is a marked symmetry between distributions of ranking vectors based on ingoing links and those of vectors based on outgoing links.  

\begin{figure}[h]
\begin{center}
\includegraphics*[width=7.cm]{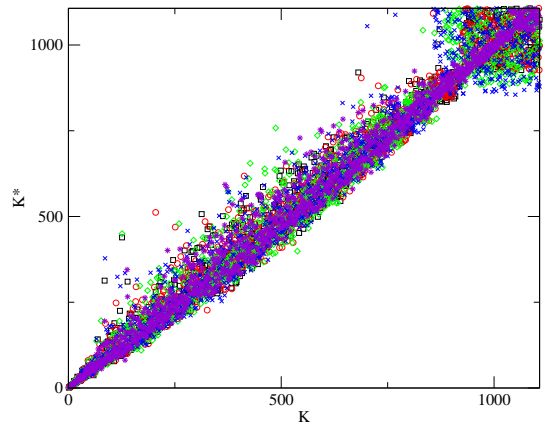}
\end{center}
\caption{(Color online) K* vs K where K (resp. K*) is the rank of a vertex when ordered  according to PageRank vector (resp CheiRank) for Honinbo (black squares), Meijin (red circles), Judan (green diamonds), Kisei (blue crosses) and amateur (violet stars) databases. 
\label{ranks}}
\end{figure}

In order to further shed light on this symmetry, we plot in Fig.~\ref{ranks} the correlation between the PageRank and the CheiRank for the five databases considered. In all these cases, there is a remarkably strong correlation between these rankings based respectively upon ingoing and outgoing links. In the WWW, there is a difference of nature between ingoing and outgoing links: webmasters are free to create as many outgoing links as they wish from their webpage, whereas the ingoing links depend on the cumulative effect of all other webmaster behaviors. In contrast, for the go network, the fact that a link is ingoing or outgoing depends on the chronological order in which the moves are played. The results displayed in Fig.~\ref{ranks} thus seem to indicate that there is a strong correlation between moves which open many possibilities of new moves
and moves that can follow many other moves. However, the symmetry is
far from exact, as can be seen in Fig.~\ref{ranks}. 

Although there is always some correlation between the different ranking vectors, they usually can be quite different, for example in Wikipedia articles  \cite{dima}. A recent analysis of the world trade network \cite{trade} showed such a symmetry when all commodities were aggregated, but the symmetry was much less visible when each different commodity was treated separately. It is possible that in our case the symmetry is made more prominent by our definition of  moves through
$3\times 3$ plaquettes. A more refined approach with larger plaquettes may thus disambiguate the moves and give different results.  We nevertheless think that our result indicate a specific feature of the game, such as the existence of many short sequences of tactical moves which can be played at different moments of the game.

\section{Eigenvectors of the Google matrix}

\begin{figure}[h]
\begin{center}
\includegraphics*[width=8.cm]{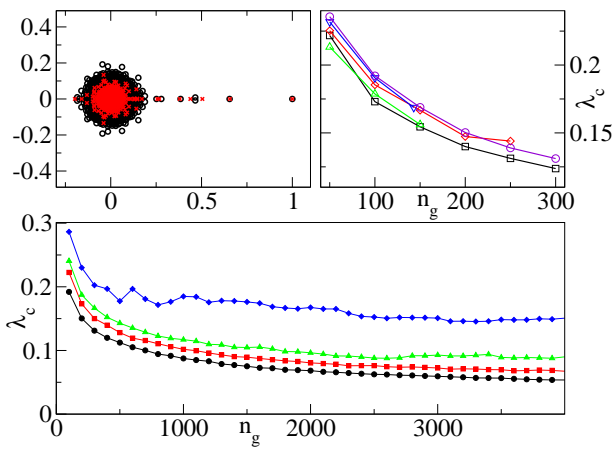}
\end{center}
\caption{(Color online) Top left:  eigenvalues in the complex plane for matrices
$G$, $d=4$, $\alpha=1$; black circles: Honinbo; red crosses: amateur.
 Bottom: $\lambda_c$ such that from top to bottom $99\%$, $95\%$, $90\%$,
$80\%$ of eigenvalues $\lambda$ 
verify $|\lambda| <\lambda_c$ for amateur games. Top right: $\lambda_c$ for $80\%$ of eigenvalues for our 5 databases, same color code as in Fig.~\ref{zipf_all}.\label{nuagesvp}}
\end{figure}

\begin{figure}[hbt]
\begin{center}
\includegraphics*[width=7.5cm]{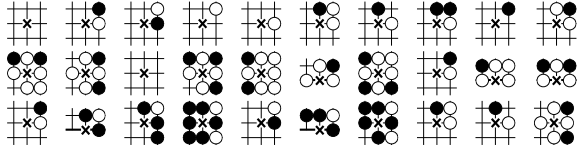}
\end{center}
\caption{Moves corresponding to the 10 largest entries of right eigenvectors of $G$ for eigenvalues $\lambda_1$ (PageRank)(top), $\lambda_3$ (middle) and $\lambda_7$ (bottom), Honinbo database. Black is playing at the cross. Top line coincides with the 10 most frequent moves. \label{frequent_moves}}
\end{figure}

As can be seen in Fig.~\ref{log_sorted_pageranks}, the ranking
vectors are distributed according to power laws and thus are mainly localized on few vertices, mainly the most frequent ones according to Zipf's law (see e.~g.~Fig.~\ref{frequent_moves}, upper line). 
%All vectors are localized on a few sites, the main one being the empty plaquette. 
However, these ranking vectors correspond to the eigenvector associated with the largest eigenvalue of different matrices built from the network. We now consider the other eigenvectors of $G$.  In particular, the eigenvectors associated with next to leading eigenvalues can describe specific communities inside the network \cite{google}.  The distribution of eigenvalues is also
important, as it reflects the structure of the network \cite{google}.
%To further refine the analysis, we take into account other eigenvalues. 
Fig.~\ref{nuagesvp} shows the complex eigenvalues of the matrix $G$. For the WWW a sizable 
fraction of eigenvalues are close to zero, while the remaining ones are 
spread inside the unit circle, with no gap between the first eigenvalue 
and the bulk \cite{google}. By contrast, in the case of the go network, there is a 
huge gap between the first eigenvalue $\lambda=1$ and the next ones.   This is reminiscent of what can be seen in some lexical 
networks~\cite{google}.  Such features 
indicate that the network is well-connected, with few isolated 
communities, and is consistent with the finding of a high clustering coefficient
(see inset of Fig.~\ref{log_sorted_pageranks}). Whereas the WWW contains many communities of webpages which are almost cut off from the rest of the network, this is not the case
for the go network, where communities -- i.e. sequences of tactical moves preferentially played together -- have more connections to the rest of the network, indicating that tactical moves can belong to different strategic groupings. To put these data in perspective, we have constructed a random version of the network, by randomly shuffling the moves inside each game of the databases. This process conserves Zipf's law and the global characteristics, but eigenvalues of $G$ are 
all concentrated in the bulk (data not shown), in contrast with the real go network, indicating that communities are destroyed by the randomization process.

\begin{figure}[h]
\begin{center}
\includegraphics*[width=8.cm]{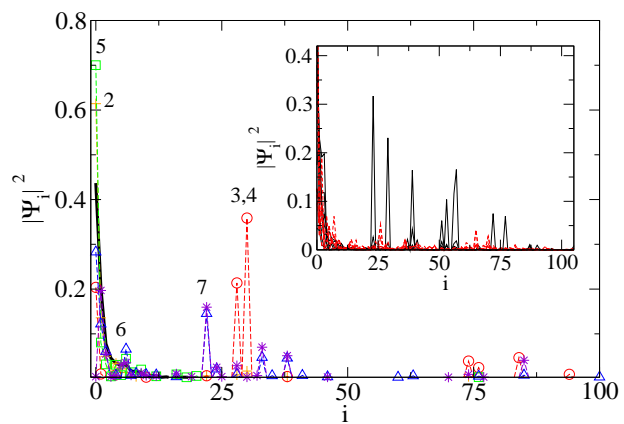}
\end{center}
\caption{(Color online) Moduli squared of the right eigenvectors associated with the 7 largest eigenvalues $|\lambda_1|=1>|\lambda_2|...>|\lambda_7|$ of $G$ (Honinbo database) for the first 100 moves in 
decreasing frequency ($|\lambda_1|$ (PageRank): thick black line, $|\lambda_2|$: orange pluses; $|\lambda_3|=|\lambda_4|$: red circles;  $|\lambda_5|$: green squares;  $|\lambda_6|$: blue triangles; $|\lambda_7|$: violet stars). Inset:
Same for amateur database (full black line) and random network
(dashed red line, see text). \label{eigenv}}
\end{figure}

 In Fig.~\ref{nuagesvp} (bottom), it is shown that  the radius of the bulk of eigenvalues changes with the number of games $n_g$ entered in the network. This indicates that as more games are taken into account, rare links appear which break more and more the weakly coupled communities. 
%The fact that the spectral radius is smaller for professional is probably due to the fact that professional are more prone to play rare sequences than amateurs. 
The next to leading eigenvalues are important, as they indicate the presence of communities of moves which have common features. The distribution of the first 7 eigenvectors (Fig.~\ref{eigenv}) shows that they are concentrated on particular sets of moves
different for each vector.  The corresponding moves are displayed 
in Fig.~\ref{frequent_moves} for the Honinbo database. The first eigenvector is mainly localized on the most frequent moves. By contrast, the third one is localized on moves describing captures of the opponent's stones, and part of them single out  the well-known situation of {\em ko} (``eternity''), where players repeat captures alternately. The 7th eigenvector singles out moves which appear to protect an isolated stone by connecting it with a chain.
These eigenvectors are different for different tournaments and from professional to amateur. Indeed, the 
inset of Fig.~\ref{frequent_moves} shows the distribution of the first seven eigenvectors for amateur database, very different from the one for Honinbo. It also shows for comparison the distribution for the randomized network (see above), which is much less peaked.
Systematic studies of these eigenvectors, 
 as well as the frequency of sequences of moves,
should enable to group together certain moves, 
and should help to elaborate efficient go simulators.

%%%%%%%%%%%%%%%%%%%%%%%%%%%%%%%%%%%%%%%%%%%%%%%%%%%%%%%%%%%%%%%%%%%%%%%%%%%%%%%
%******************************************************************************
\section{Conclusion}
%******************************************************************************
%%%%%%%%%%%%%%%%%%%%%%%%%%%%%%%%%%%%%%%%%%%%%%%%%%%%%%%%%%%%%%%%%%%%%%%%%%%%%%%

In this paper, we have studied the game of go, one of the most ancient and
complex board games, from a complex network perspective.  
%We have shown that
%taking 
%into account the local environment around each move 
%creates a hierarchization among moves.  
We have defined
a proper categorization of moves taking 
into account the local environment, and shown that in this
case Zipf's law emerges from data taken from different tournaments.
The network of go moves has some peculiarities, such
as a statistical symmetry between ingoing and outgoing links distributions, which reflects
itself in a symmetry between rankings based on ingoing on outgoing links, a feature not
seen in many other complex directed networks such as the WWW. 
Differences between professional tournaments and amateur games can be seen.
Properties of eigenvalues and eigenvectors of the matrices producing
ranking vectors vary between amateur and different professional tournaments. 
Certain eigenvectors are localized on specific groups of moves which correspond
to different strategies.  We think that the point of view developed
in this paper should allow to better modelize such games and could also help to
design simulators which could in the future beat good human players.  Our approach could be used for other types of games, and in parallel shed light on the human decision making process.

We thank CalMiP for the use of their supercomputers, and D. Shepelyansky for useful discussions.\\

%%%%%%%%%%%%%%%%%%%%%%%%%%%%%%%%%%%%%%%%%%%%%%%%%%%%%%%%%%%%%%%%%%%%%%%%%%%%%%%
%******************************************************************************

\end{document}